\documentclass[final,1p,times]{elsarticle} 
\usepackage{graphicx}
\usepackage{amssymb} 
\usepackage{amsthm} 
\usepackage{lineno}
\usepackage{comment}
\usepackage{subfigure}
\usepackage{bbm}
\usepackage{amsmath}
\usepackage{amsfonts,amsbsy}

\newcommand{\gc}{{\gamma-\mathrm{ch}}}

\newcommand{\la}{\left<}
\newcommand{\ra}{\right>}
\newcommand{\lf}{\left(}
\newcommand{\rf}{\right)}

\newcommand{\ph}{\gamma}
\newcommand{\ch}{\mathrm{ch}}

\newcommand{\nc}{\la N_\ch \ra}
\newcommand{\np}{\la N_\ph \ra}

\newcommand{\be}{\begin{equation}}
\newcommand{\ee}{\end{equation}}
\newcommand{\bea}{\begin{eqnarray}}
\newcommand{\eea}{\end{eqnarray}}
\newcommand{\ndyn}{\nu_{\mathrm{dyn}}}

\journal{Nuclear Physics A} 

\begin{document}

%\linenumbers
\begin{frontmatter} 

% Your Title - please insert
\title{Search for QCD Phase Transitions and the Critical Point Utilizing Particle Ratio Fluctuations and Transverse
Momentum Correlations from the STAR Experiment}

%% Single author (and collaboration) - please insert
\author{Prithwish Tribedy (for the STAR Collaboration)}
\address{Variable Energy Cyclotron Centre, 1/AF Bidhan Nagar, Kolkata-700064, India}
%
%% Multiple authors
%\author[auth2]{Marcus Junius Brutus}
%\address[auth1]{Somewhere, Rome}
%\address[auth2]{Somewhere else, Rome}

\begin{abstract} 
Dynamical fluctuations of the globally conserved quantities in heavy ion collision such as baryon number, strangeness, charge, and isospin are suggested to carry information about the deconfinement and chiral phase transitions.
The STAR experiment has performed a comprehensive study of the collision energy and charge dependence of dynamical particle ratio ($K/\pi$, $p/\pi$, and $K/p$) fluctuations, net-charge fluctuations, and transverse momentum correlations %in the STAR TPC
at mid-rapidity, as well as neutral-charge pion fluctuations at forward rapidity. 
The centrality, charge, and collision energy dependence from new measurements of the fluctuation observables $\ndyn$ and $r_{m,1}$ and the energy dependence of transverse momentum correlations from $\sqrt{s_{NN}}$ = 7.7-200 GeV Au+Au collisions are presented. These results are also compared to predictions from hadronic models.

\end{abstract} 
\end{frontmatter} 
\section{Introduction}
Event by event fluctuations in production of the globally conserved quantities and their correlation have always been important probes in the context of Quantum Chromodynamics (QCD) phase transition \cite{Jeon, Corrreview}. 
The goal is to search for any non-monotonic behaviour of corresponding observables as a function of collision energy which might indicate possible signature of critical point. 
Anomalous fluctuation in the production of pions of different isospins is predicted to occur for systems going through the QCD chiral phase transition \cite{dcc}. Such phenomena like formation of domains of Disoriented Chiral Condensate (DCC) would correspond to deviation in neutral-charge correlation from generic expectation. Slopes of the transverse momentum spectra are related to the temperatures of the system. Two particle transverse momentum correlation would thus be a measure of temperature fluctuation \cite{Corrreview, ptcorr}. 
 In recent Beam Energy Scan (BES) program of RHIC, the STAR experiment has collected data for Au+Au collisions at $\sqrt{s_{NN}}$ = 7.7, 11.5, 19.6, 27, 39, 62.4, and 200 GeV. 
 The observables studied are dynamical fluctuations of particle ratio ($K/\pi$, $p/\pi$, $K/p$, $\mathrm{ch^{+}/ch^{-}}$, $\gamma/\mathrm{charge}$) 
and two particle transverse momentum correlation. 
\section{Experiment and Observables}
Measurement of identified and inclusive charged particles are done at mid-rapidity ($|\eta|<1$) using a combination of Time Projection Chamber (TPC) and Time of Flight (TOF) detectors. A Photon Multiplicity Detector (PMD) and Forward Time Projection Chamber (FTPC) are used to measure event-by-event photon and charged particle multiplicity over a common rapidity range of $-3.7<\eta<-2.8$. Details for particle identification relevant to current studies can be found in Ref.\cite{kpi, terry, dogra}. Centrality selections are done using raw charged track multiplicity from the TPC.
In this analysis, three observables are used;  (i) ratio fluctuation measure is defined \cite{nudyn} as 
\be
\nu_{\mathrm{dyn}}^{M-N}  =\frac{\la M(M-1)\ra}{\la M^2\ra} + \frac{\la N(N-1)\ra}{\la N^2\ra} - 2 \frac{\la M N \ra}{\la M \ra \la N\ra},
\ee
where $M$ and $N$ are multiplicities of two particle species ($K,\pi, p, \ph, \mathrm{charge}^{\pm}$, total charge) for which ratio fluctuation is considered. The event average is defined by $<>$. By design $\ndyn$ is zero for Poissonian fluctuation. (ii) Additionally one more observable has been used as a measure of $\gc$ correlation. As introduced by Mini-Max \cite{minimax} collaboration, the observable
\be
 r_{m,1}^\gc=\frac{\la N_\ch(N_\ch-1) \cdots \lf N_\ch-m+1\rf \, N_\ph \ra \nc}{\la N_\ch(N_\ch-1) \cdots (N_\ch-m)\ra \np},
\label{eq_rm1}
\ee
 for generic pion production scenario (and for Poissonian fluctuation) by design gives a value of unity for all its order $m$ and higher order becomes increasingly sensitive to any form of correlation or anti-correlation \cite{minimax,dccmodel}. For DCC like scenario this observable will become $1/(m+1)$. (iii) Observable for the two particle momentum correlation is defined \cite{ptcorr} as
\bea
\la \Delta p_{t,i}  \Delta p_{t,j} \ra = \frac{1}{N_{event}} \sum \limits_{k=1}^{N_{event}} \frac{C_k}{N_k (N_k -1)}  \, \,   ,  \, \, 
C_k = \sum \limits_{i=1}^{N_k} \sum \limits_{j=1,i\ne j}^{N_k} \left( p_{t,i} - \la \la p_t \ra \ra \right ) \left( p_{t,j} - \la \la p_t \ra \ra \right)
\eea
where the double average $\la \la p_t \ra \ra $ is obtained by first averaging over all particles in an event and then averaging over all events considered.
All the observables defined above, by construction do not have explicit efficiency dependence.
\begin{figure}[t]
\begin{center}
\subfigure[]{
\label{fig:kp}
\includegraphics[width=0.4\textwidth]{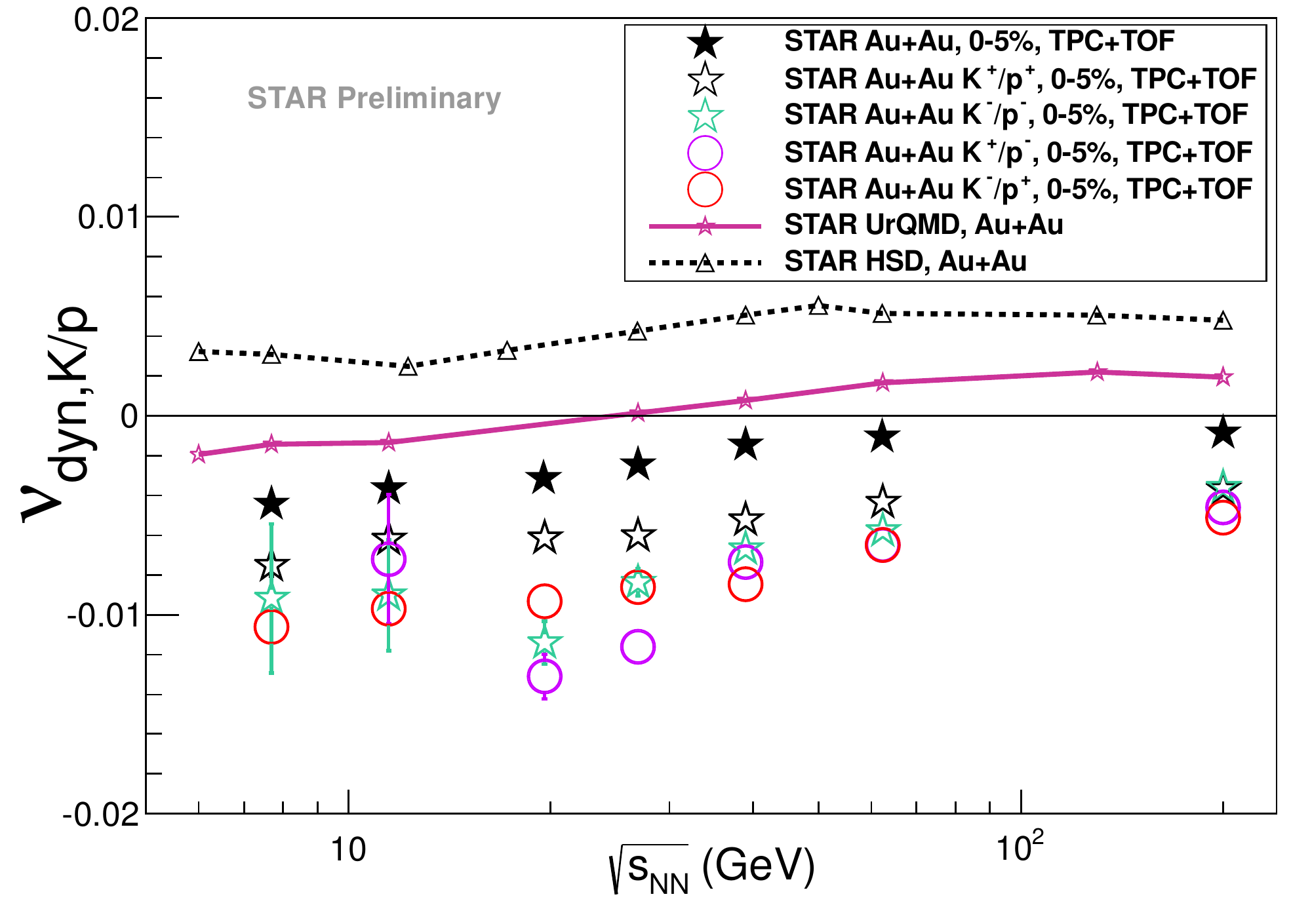}}
\subfigure[]{
\label{fig:ppi}
\includegraphics[width=0.4\textwidth]{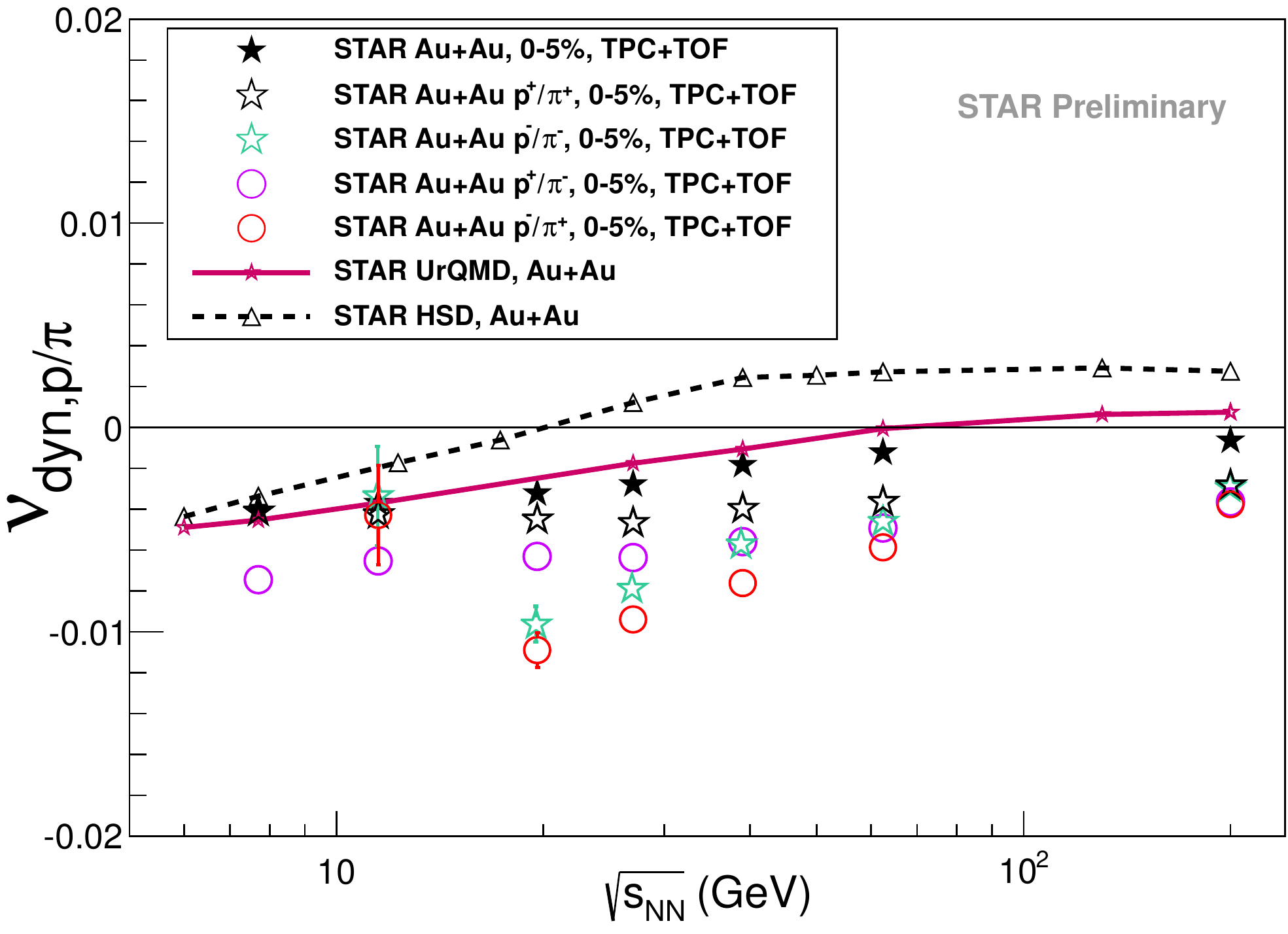}}
\subfigure[]{
\label{fig:kpi}
\includegraphics[width=0.4\textwidth]{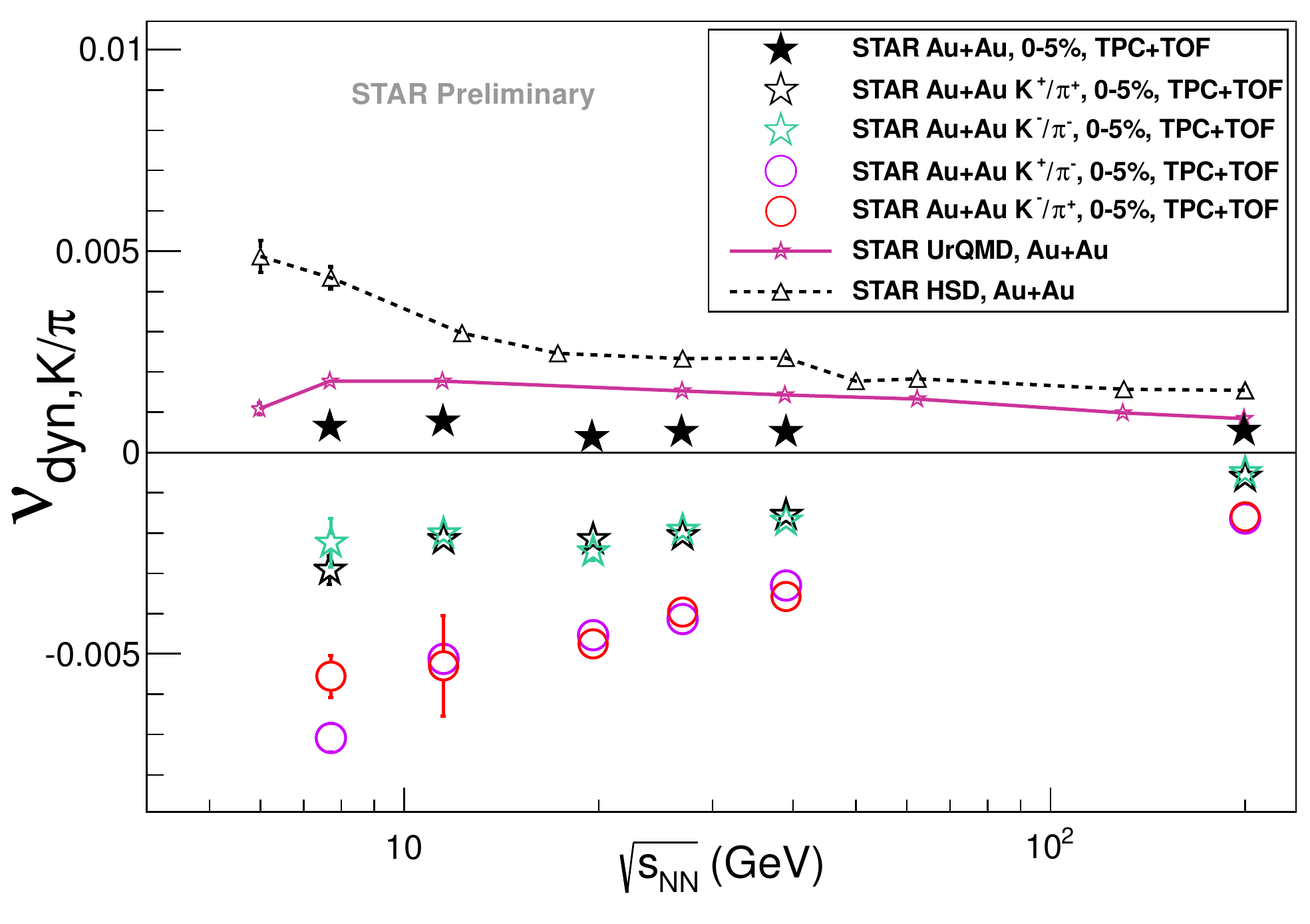}}
\subfigure[]{
\label{fig:chpm}
\includegraphics[width=0.4\textwidth]{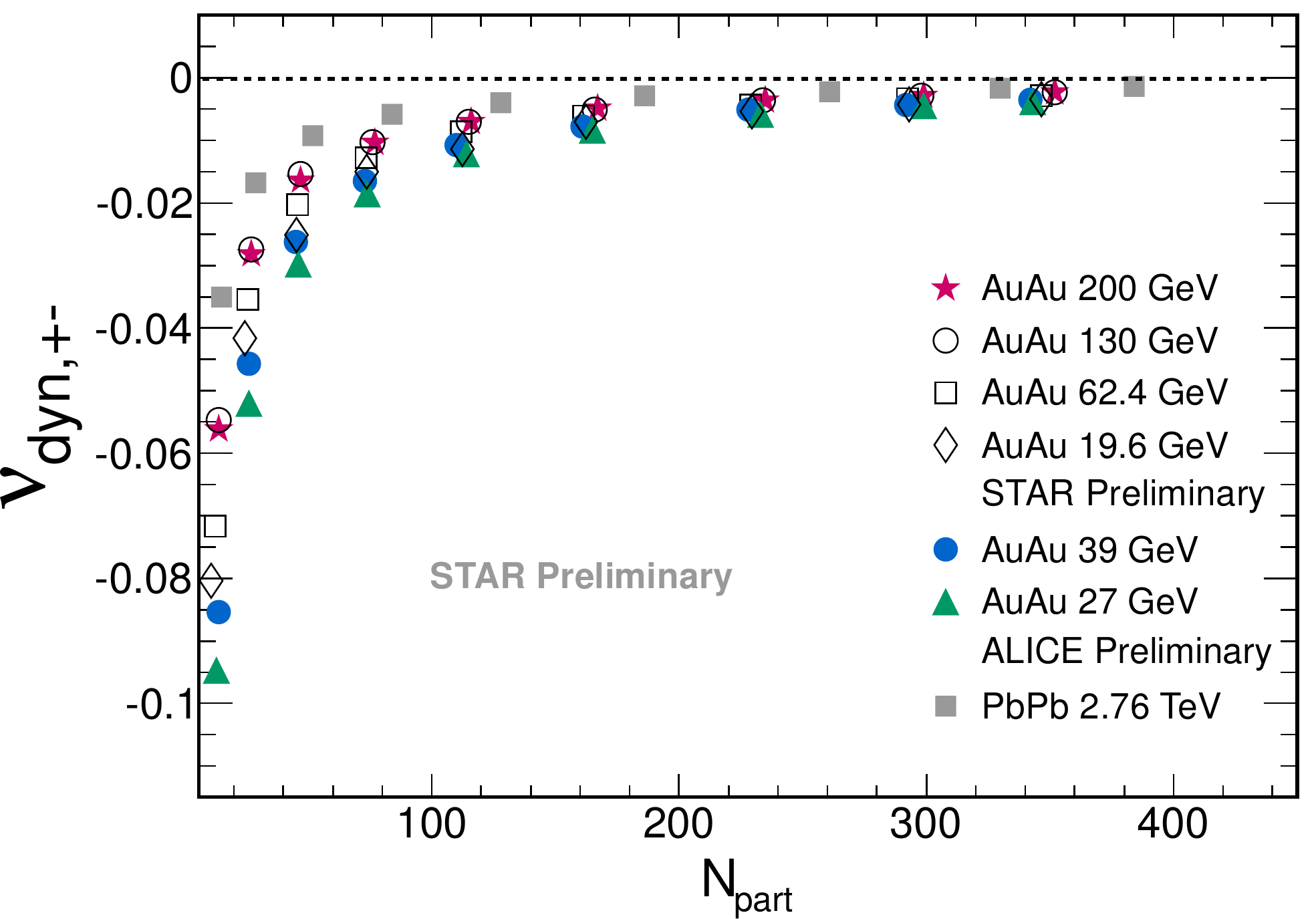}}
\caption{(color online) Observable $\ndyn$ for identified particles(a-c) and inclusive charged particles(d) at mid-rapidity for BES. Symbols represent data points and model calculations are shown by lines. Errors bars are statistical.}
\end{center}
\end{figure}
\section{Results and discussion}
\begin{figure}[t]
\begin{center}
\subfigure[]{
\label{fig:ndyn_chph}
\includegraphics[width=0.395\textwidth]{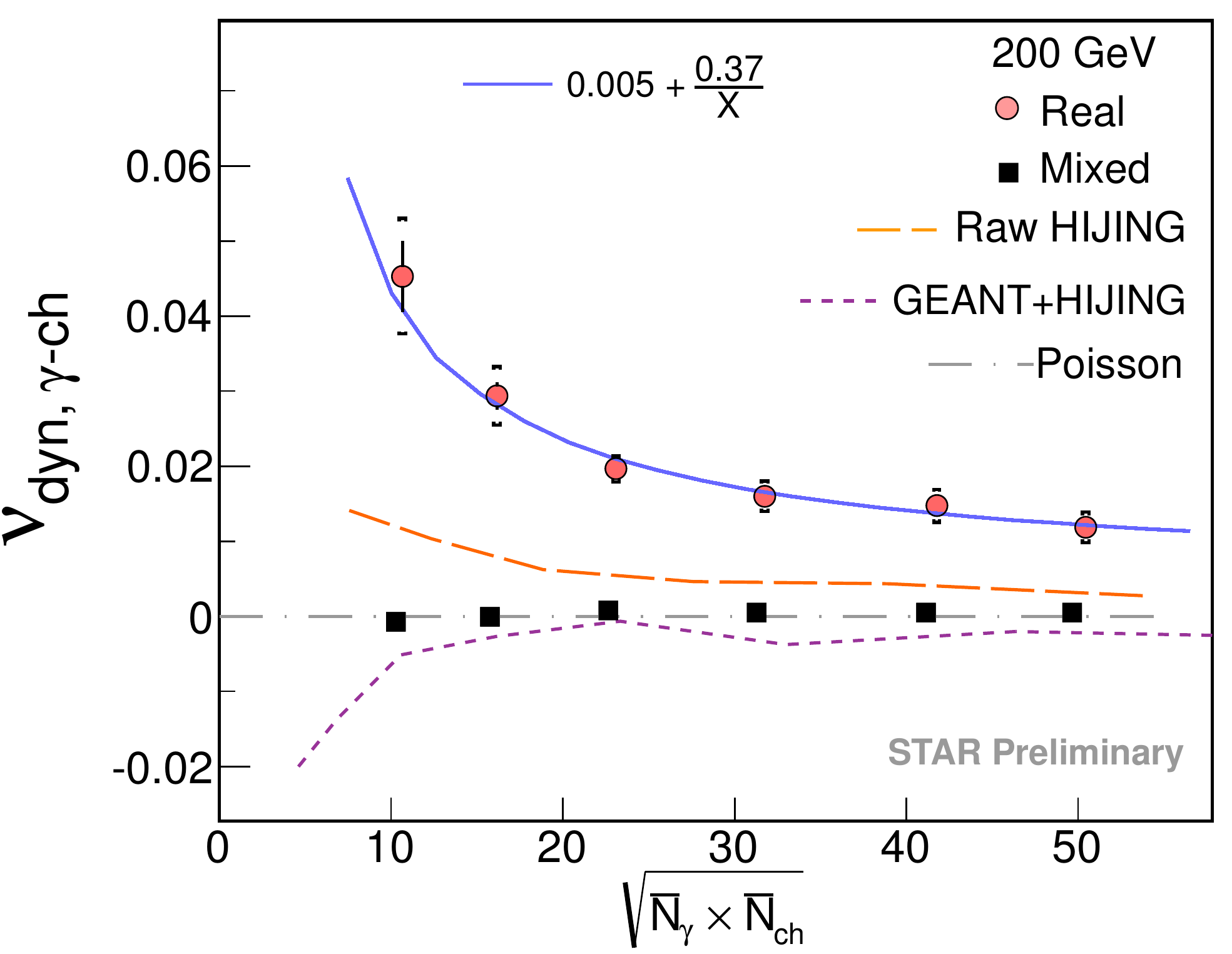}}
\subfigure[]{
\label{fig:rm1_200}
\includegraphics[width=0.395\textwidth]{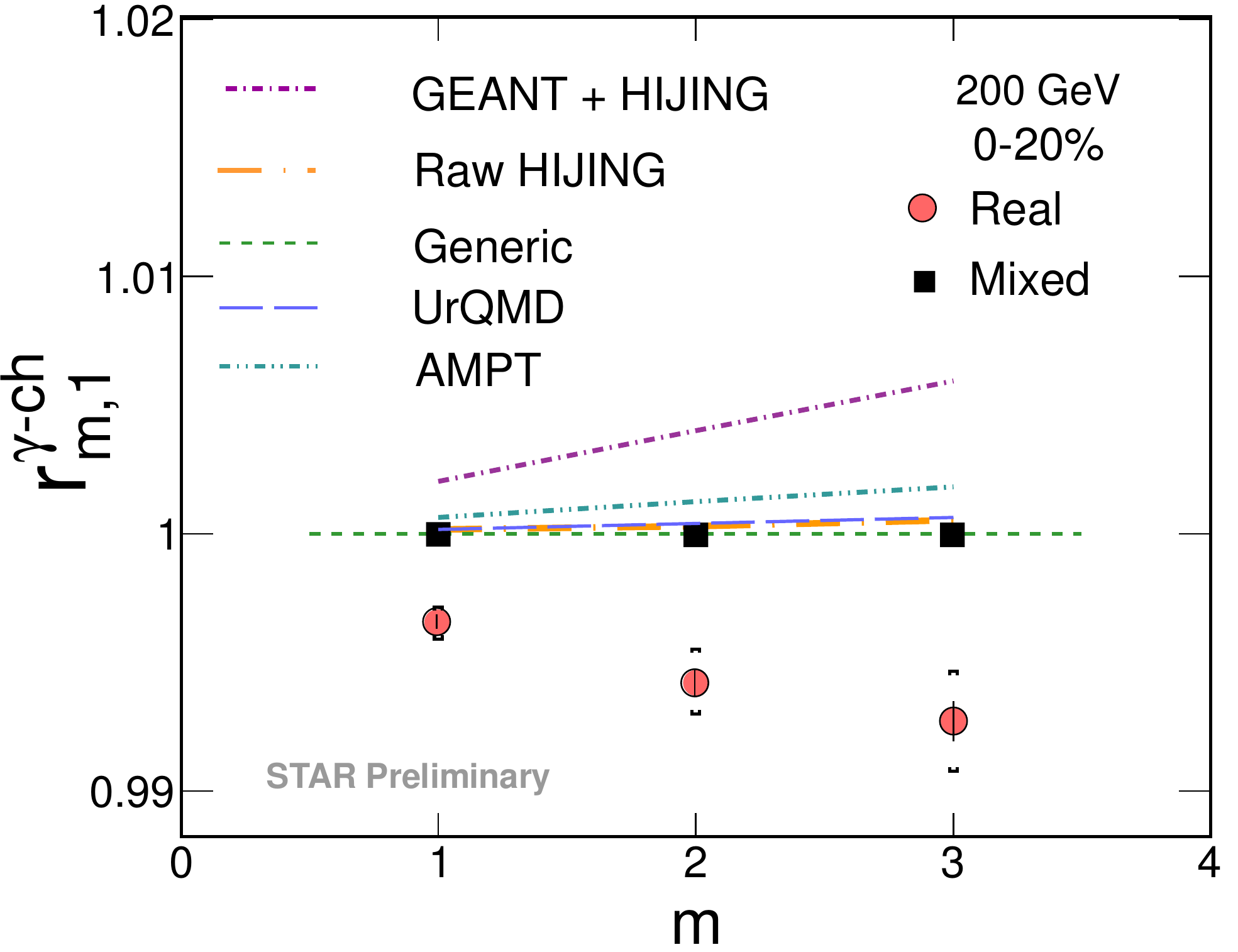}}
\subfigure[]{
\label{fig:rm1_all}
\includegraphics[width=0.385\textwidth]{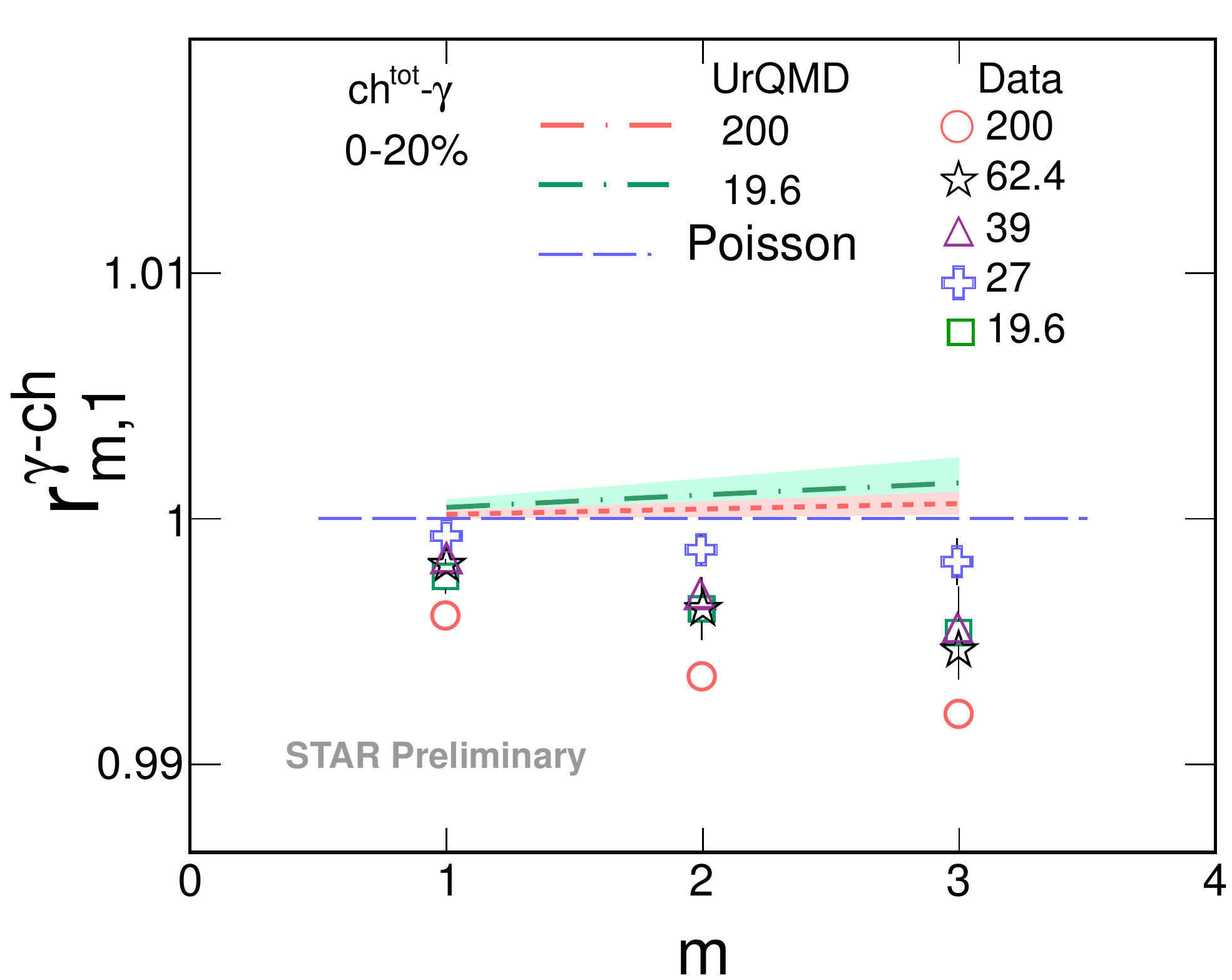}}
\subfigure[]{
\label{fig:ptcorr}
\includegraphics[width=0.395\textwidth]{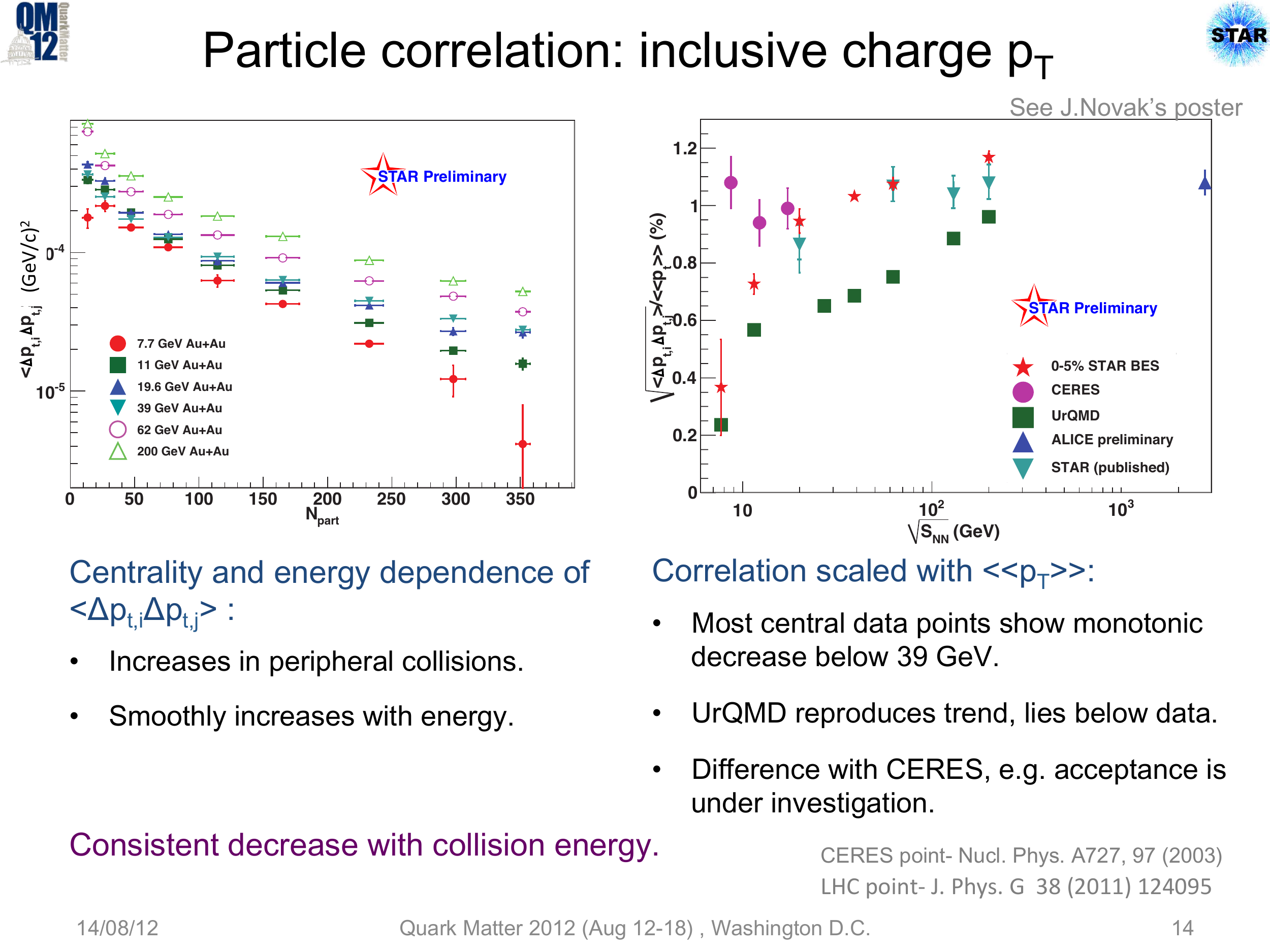}}
\caption{(color online) Observable $\nu_{dyn}$ and $r_{m,1}$ for $\gc$ correlation at 200 GeV (a-b). Markers show data points and lines are model calculations. Errorbars are both systematic (bars) and statistical (lines). (c)Energy dependence of $r_{m,1}$. (d) Energy dependence of scaled $p_T$ correlation, star markers show new measurements from STAR. Errors are statistical.}
\end{center}
\end{figure}

Excitation functions for the fluctuation measure $\ndyn$ of $K/p$, $p/\pi$ and $K/\pi$ ratios and their charge dependence over an energy range of $\sqrt{s_{NN}}$ = 7.7-200 GeV are shown in (Fig.\ref{fig:kp}-\ref{fig:kpi}). In all the cases two lines represent model calculations and solid makers show $\ndyn$ for ratio of all charges in the species considered for most central ($0-5\%$) events. For both $K/p$ and $p/\pi$ we see the observable $\ndyn$ is negative which indicates that the production of corresponding pairs are highly correlated. We see either a weak energy dependence or monotonic decrease with decreasing energy. Hadronic models like $\mathrm{HSD}$ and UrQMD do not seem to follow the sign of the data, only trend is qualitatively reproduced. All four combinations of charge dependent excitation functions show negative value of $\ndyn$ indicating correlated production from decays of resonances like $\Delta^{++}, \Delta^{0}, \Lambda^0, \Omega$. Particularly for $K/\pi$ we can see that the opposite-sign pairs are more correlated compared to like-sign pairs indicating dominance of neutral resonance decays. The charge dependence of the excitation function begin to dominate at lower energy. The fluctuation of positive-negative charge particle ratio is always dominated by decay of several neutral resonances like $\rho^0, K_S, \eta^{\prime}, \omega$ resulting negative values of observable $\ndyn$ as shown in Fig. \ref{fig:chpm}. The data presented here are measurement at mid-rapidity and consistent with a similar measurement done at forward rapidity.  The other interesting observation is that $\ndyn$ follows an approximate $1/N_{\mathrm{part}}$ scaling at all energies. The similar scaling is observed with average multiplicity for $\gc$ ratio fluctuation as shown in Fig.\ref{fig:ndyn_chph} where data for 200 GeV is compared to mixed event and model calculations. Systematic errors due to different kinematic cuts and variation in collision vertex position have been included in these results; estimation of further systematic uncertainties due to charge particle contamination contribution are underway. Results for most central ($0-20\%$) events using the observable $r_{m,1}$ is shown in Fig.\ref{fig:rm1_200}-\ref{fig:rm1_all}. Data seem to show a different trend from mixed events and several conventional model calculations at all energies. Even though data show difference from generic (Poisson) expectation and conventional models, the corresponding physical origin requires further systematic investigation. The results for new measurements of $p_T$ correlation at mid-rapidity is shown in Fig.\ref{fig:ptcorr}. Over the energy range 39-200 GeV data show a flat nature of the excitation function of scaled correlation $\la \Delta p_{t,i}  \Delta p_{t,j} \ra /  \la \la p_T \ra \ra$ for most central events in contrast to UrQMD model calculation. Below 39 GeV, data seem to show a decreasing trend in contrast to previous measurements done by CERES collaboration. The difference could be due to acceptance difference which is under investigation.%$\!\!$
\section{Summary}
We have studied ratio fluctuation of various particle species. For $p/\pi$ and $K/p$ we see monotonic behavior with collision energy dominated by correlated productions. No strong energy dependence was observed for $K/\pi$ ratio. The charge dependence of particle ratio fluctuations exhibit differences between same and opposite sign pairs, becoming dominant at low energy. The inclusive positive-negative charge ratio fluctuation shows monotonic decrease. The $\gc$ correlation result shows difference from conventional models in the energy range 19.6 -200 GeV. The two particle transverse momentum correlation result shows a weak energy dependence above 39 GeV up to 2.76 TeV but decreases with incident energy below 39 GeV. In summary, results of ratio fluctuations and correlation measured in the energy range (7.7 - 200 GeV) do not show any non-monotonic trend and cannot be fully explained by hadronic models.
%
%
%\section{References}
%


\begin{thebibliography}{00} 
\bibitem{Jeon} V. Koch, arXiv:0810.2520 [nucl-th] (2008); S.~Jeon and V.~Koch, Phys.\ Rev.\ Lett.\  {\bf 83}, 5435 (1999).

\bibitem{Corrreview} H. Heiselberg, Phys. Rep. 351, 161 (2001).

\bibitem{dcc} J.D. Bjorken, What lies ahead?, SLAC-PUB-5673, 1991; J.~P.~Blaizot and A.~Krzywicki, Phys.\ Rev.\  D {\bf 46}, 246 (1992);  K.~Rajagopal and F.~Wilczek, Nucl.\ Phys.\  B {\bf 399}, 395 (1993); K.~Rajagopal, arXiv:hep-ph/9504310.


\bibitem{ptcorr} J.~Adams {\it et al.} [STAR Collaboration], Phys.\ Rev.\ C {\bf 72}, 044902 (2005).

\bibitem{nudyn}
  C.~Pruneau, S.~Gavin and S.~Voloshin,
  %``Methods for the study of particle production fluctuations,''
  Phys.\ Rev.\  C {\bf 66}, 044904 (2002)
  [arXiv:nucl-ex/0204011].

%[7] B. I. Abelev et al. (STAR Collaboration), Phys. Rev. Lett. 103, 092301 (2009).

\bibitem{kpi} B. I. Abelev {\it et al.} [STAR Collaboration], Phys. Rev. Lett. 103, 092301 (2009).

\bibitem{terry} T. Tarnowsky, arXiv:1101.3351 [nucl-ex], arXiv:1106.6110v1 [nucl-ex] (2011).

\bibitem{dogra} S M Dogra, J. Phys. G: Nucl. Part. Phys. 35 (2008) 104094


\bibitem{minimax}
  T.~C.~Brooks {\it et al.}  [MiniMax Collaboration],
  %``A Search for disoriented chiral condensate at the Fermilab Tevatron,''
  Phys.\ Rev.\  D {\bf 61}, 032003 (2000)
  [arXiv:hep-ex/9906026].
  
\bibitem{dccmodel} P.~Tribedy {\it et al.} Phys.\ Rev.\  C {\bf 85}, 024902 (2012)  [ arXiv:1108.2495 [nucl-ex]].


  
\end{thebibliography}
\end{document}